\documentclass[12pt]{iopart}

\usepackage{graphicx}

\renewcommand{\d}{{\rm d}}

\begin{document}

\title[Mean field theory for a balanced hypercolumn model in V1]{Mean
field theory for a balanced hypercolumn model of orientation
selectivity in primary visual cortex}

\author{A Lerchner\dag, G Sterner\ddag, J Hertz\S\ and M Ahmadi\S}

\address{\dag\ {\O}rsted-DTU, Technical University of Denmark, 2800 Kgs. Lyngby,
Denmark}
\address{\ddag\ Department of Physics, University of Rochester, Rochester NY
14627}
\address{\S\ Nordita, Blegdamsvej 17, 2100 Copenhagen {\O}, Denmark}

\ead{al@oersted.dtu.dk}

\begin{abstract}
We present a complete mean field theory for a balanced state of a
simple model of an orientation hypercolumn. The theory is
complemented by a description of a numerical procedure for solving
the mean-field equations quantitatively. With our treatment, we
can determine self-consistently both the firing rates and the
firing correlations, without being restricted to specific neuron
models. Here, we solve the analytically derived mean-field
equations numerically for integrate-and-fire neurons. Several
known key properties of orientation selective cortical neurons
emerge naturally from the description: Irregular firing with
statistics close to -- but not restricted to -- Poisson
statistics; an almost linear gain function (firing frequency as a
function of stimulus contrast) of the neurons within the network;
and a contrast-invariant tuning width of the neuronal firing. We
find that the irregularity in firing depends sensitively on
synaptic strengths. If Fano factors are bigger than 1, then they
are so for all stimulus orientations that elicit firing. We also
find that the tuning of the noise in the input current is the same
as the tuning of the external input, while that for the mean input
current depends on both the external input and the intracortical
connectivity.
\end{abstract}

\submitto{Network: Computation in Neural Systems}


\section{Introduction}

Neurons in primary visual cortex (V1) fire highly irregularly in
response to visual stimuli, but with reproducible firing rates.
They do so despite the fact that they receive synaptic input from
thousands of other cortical neurons, which would lead to
fluctuations in the input that were small compared to the mean if
excitatory and inhibitory inputs were not balanced
\cite{Softky+Koch:1993}. There has been some success in describing
how such a balance can emerge self-consistently from dynamics that
are plausible for cortical networks. This was accomplished by mean
field-descriptions by van Vreeswijk and Sompolinsky
\cite{vVreeswijk+Sompolinsky:1996, vVreeswijk+Sompolinsky:1998}
and Amit and Brunel \cite{Amit+Brunel:1997a, Amit+Brunel:1997b,
Brunel:2000}. However, their treatments do not permit a
self-consistent calculation of firing correlations. How to do this
correctly was first shown for an all-inhibitory network by Hertz
\emph{et al.}~\cite{Hertz+Richmond+Nilsen:2003} using the
systematic formulation of mean field theory due to Fulvi Mari
\cite{FulviMari:2000}. In a recent paper \cite{Lerchneretal:2004}
we presented a mean-field theory for a balanced network model that
allowed us to quantify how the irregularity in firing and, more
generally, the firing correlations depend on intrinsic network
properties such as synaptic strengths. The analysis was applied to
a statistically homogeneous network, representing a cortical
column composed of neurons with similar response characteristics.
Here, we show how to extend this treatment to networks with
systematic structure, consisting of multiple cortical columns. In
particular, we model an orientation hypercolumn, composed of a set
of orientation columns.

An orientation column contains neurons that respond strongest to
elongated visual stimuli of a specific orientation, the
\emph{preferred orientation} (PO). Orientation selective neurons
exhibit a tuned response to other orientations, with sharply
decreasing firing rates as the similarity between PO and stimulus
orientation decreases, until the firing is completely suppressed
for orientations outside the \emph{tuning width} of the neuron in
question. An important feature of orientation tuning is that the
tuning width is independent of the stimulus contrast
\cite{Sclar+Freeman:1982}. It is not possible to capture this
feature in a single-neuron description using a Hubel and Wiesel
feed-forward connectivity \cite{Hubel+Wiesel:1962} from the
lateral geniculate nucleus (LGN); rather, cortical interactions
are needed to achieve contrast-invariant tuning (for review see
\cite{Sompolinsky+Shapley:1997}). Ben-Yishai \etal
\cite{Ben-Yishai+LevBar-Or+Sompolinsky:1995} proposed a model for
which the tuning width is independent of the contrast, but a
threshold-linear relationship between input current and firing
rate was an assumption of the model, and the problem of the firing
statistics was not addressed.

Here, we show how a contrast-invariant tuning width, an almost
linear input-output relationship, and irregular firing can all be
explained by a balanced hypercolumn model. With our mean-field
treatment, we can quantify how certain network properties like
synaptic strengths, tuning of the LGN input and of the
intracortical connectivity influence the statistics and tuning of
the neuronal firing. Using the Fano factor $F$ (the ratio of spike
count variance and mean spike count) to quantify the irregularity
in firing, we find, e.g., that if $F$ is significantly greater
than 1 the orientation tuning of $F$ reaches a maximum at the PO
(Fano factors greater than 1 are normally observed for neurons in
V1 \cite{Gershon+etal:1998}). We also make quantitative
predictions about the tuning of the input currents and their
fluctuations.

\section{Model and Methods}
\label{sec:model}

\begin{figure}[t]
\centering
 \includegraphics[width=12cm]{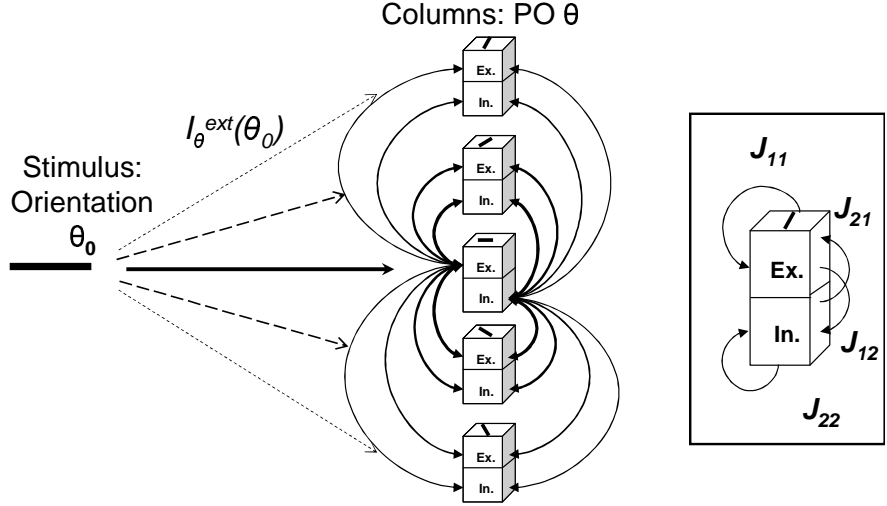}
\caption{Structure of the model network. The hypercolumn consists
of multiple orientation columns, each of which has an excitatory
and an inhibitory subpopulation and is assigned a preferred
orientation (PO) $\theta$. Columns with more similar POs share on
average more connections than more dissimilar ones (the density of
connections is indicated only between one column and the rest, for
clarity). The network receives excitatory external input, weakly
tuned to the stimulus orientation $\theta_0$. The inset shows a
sketch for the connectivity and connection strengths $J_{ab}$
within an orientation column.} \label{fig:Network}
\end{figure}

We model a single orientation hypercolumn in primary visual
cortex, with a simplified network architecture as indicated in
Figure~\ref{fig:Network}. The network comprises an excitatory
population and an inhibitory one, of sizes $N_1$ and $N_2$,
respectively. Each population is divided into $n$ sub-populations
(orientation columns), parameterized by an angle $\theta$. The
angles, spaced equally between $-\pi/2$ and $\pi/2$, indicate the
preferred orientation (PO), to which the neurons in the
corresponding column respond strongest.

We use leaky integrate-and-fire neurons and interconnect them
randomly with a connection probability $P_{ab}(\theta-\theta')$
that depends on the similarity of the POs. The probability that a
neuron with PO $\theta$ (in population $a$) receives afferent
input from a neuron with PO $\theta'$ in population $b$ is taken
as
\begin{equation}\label{eq:ConnProb}
    P_{ab}(\theta-\theta') = \frac{K_b}{N_b} \left( 1 + \gamma \cos
2(\theta-\theta') \right),
\end{equation}
where $K_b$ is the expected overall number of inputs from neurons
in population $b$. We take the ratio $K_b/N_b$ independent of $b$,
i.e., excitatory and inhibitory neurons interconnect with the same
probability in our model. The functional form of
(\ref{eq:ConnProb}) is motivated by anatomical evidence that the
connection probability between cortical neurons decreases as their
distance increases, and by the fact that orientation columns with
similar PO tend to lie closer together on the cortical surface
than ones with dissimilar PO. We followed Ben-Yishai \emph{et
al.}~\cite{Ben-Yishai+LevBar-Or+Sompolinsky:1995} in choosing the
simplest possible form that is periodic with period $\pi$. We
assume that the degree of tuning, as measured by the parameter
$\gamma \in (0,1)$, is the same for both the inhibitory and the
excitatory population.

Each nonzero synapse from a neuron in population $b$ to one in
population $a$ is taken to have strength
\begin{equation}\label{eq:Jij}
    J_{ij}^{a \theta, b \theta'} = \frac{J_{ab}}{\sqrt{K_b}}
\end{equation}
where the parameters $J_{ab}$ are of order 1. With this scaling,
the fluctuations in the input current are also of order 1, the
same order as the distance between reset and threshold of our
model neurons (cf. van Vreeswijk and Sompolinsky
\cite{vVreeswijk+Sompolinsky:1996, vVreeswijk+Sompolinsky:1998}).

The subthreshold dynamics of the membrane potentials are given by
\begin{equation}\label{eq:model}
    \frac{\d u_i^{a \theta}(t)}{\d t} = -\frac{u_i^{a \theta}(t)}{\tau}
    + I_{a \theta}^{\rm ext}(\theta_0)
    + I_i^{a \theta, {\rm rec}}(t),
\end{equation}
where the membrane time constant $\tau$ is chosen to be the same
for all neurons. The excitatory external input from the LGN, $I_{a
\theta}^{\rm ext}(\theta_0)$, is assumed to be (weakly) tuned to
the orientation $\theta_0$ of the stimulus due to a feed-forward
connectivity from the LGN as in the classical model by Hubel and
Wiesel \cite{Hubel+Wiesel:1962}. For simplicity, we take it to be
constant in time and the same for all neurons $i$ within a column.
The functional form we use is, similar to the tuning
(\ref{eq:ConnProb}) of the intracortical connectivity,
\begin{equation}\label{eq:I-ext}
    I_{a \theta}^{\rm ext}(\theta_0) = I_a^{\rm ext} (1 + \epsilon \cos 2(\theta - \theta_0)),
\end{equation}
where $\epsilon \in (0,1)$ is the degree of tuning, which is
assumed to be the same for both populations. (The condition
$\epsilon < 1$ assures $I_{a \theta}^{\rm ext}(\theta_0)$ to be
non-negative, i.e. excitatory, for all orientations). A more
detailed model for this external input current, including temporal
fluctuations and random connectivity, was briefly described in an
overview article by Hertz \emph{et al.}
\cite{Hertz+Lerchner+Ahmadi:2004}.

The recurrent input $I_i^{a \theta, {\rm rec}}(t)$ from within the
model cortex is given by
\begin{equation}\label{eq:I-rec}
    I_i^{a \theta, {\rm rec}}(t) = \sum_{b=1}^2
    \sum_{\theta' = {\theta_1}}^{\theta_n} \sum_{j = 1}^{N_b/n}
    J_{ij}^{a \theta, b \theta'} S_j^{b \theta'}(t),
\end{equation}
where $S_j^{b \theta'}(t) = \sum_s \delta(t-t_{j \theta' b}^s)$ is
the spike train of neuron $j$ with PO $\theta'$ in population $b$.

\subsection*{Mean Field Theory}

In the following mean-field description of the orientation
hypercolumn model, we consider stationary firing only, for
simplicity. However, the formulation is general enough to allow
for non-stationary rates. We presented such a time-dependent
treatment for a balanced single-column model elsewhere
\cite{Lerchneretal:2004}.

Because of the dilute random connectivity, each neuron receives a
high number of uncorrelated inputs (we assume $K_b$ to be large,
but smaller than $N_b$). According to the central limit theorem,
the recurrent input currents given by (\ref{eq:I-rec}) can
therefore be described as Gaussian random processes. For
stationary rates, the mean input current is constant in time for
any given neuron, although the level of the mean does vary from
neuron to neuron due to the random connectivity. In a general
mean-field theory, one must consider temporal correlations in
these currents, i.e., not restrict the description of the random
processes to white noise.

To separate the mean of the currents from their fluctuations
(``noise"), it is convenient to apply such separations to the
description of both the synaptic weights $J_{ij}^{a \theta, b
\theta'}$ and the spike trains $S_j^{b \theta'}(t)$ in
(\ref{eq:I-rec}). For the weights we can write
\begin{equation}\label{eq:J-separation}
    J_{ij}^{a \theta, b \theta'} = \overline{J_{ij}^{a \theta, b \theta'}}
    + \delta J_{ij}^{a \theta, b \theta'},
\end{equation}
where the bar means averaging over the index $j$, i.e., the
neurons in the source population:
\begin{equation}\label{eq:J-bar}
    \overline{J_{ij}^{a \theta, b \theta'}} = \frac{1}{N_b/n}
    \sum_{j = 1}^{N_b/n} J_{ij}^{a \theta, b \theta'}
\end{equation}
Generally, we use the bar-notation for averaging over neuron
populations, which will always apply to the running index $j$ in
this work. To separate the spike trains into static and dynamic
components, we write
\begin{equation}\label{eq:S-separation}
    S_j^{b \theta'}(t) = r_b(\theta') + \delta r_j^{b \theta'} +
    \delta S_j^{b \theta'}(t),
\end{equation}
where $r_b(\theta') = \overline{r_j^{b \theta'}} = 1/(N_b/n)
\sum_j r_j^{b \theta'}$ is the average rate of the neurons in
sub-population $\theta'$ of population $b$. The difference between
this average rate and the actual rate of neuron $j$ is denoted
$\delta r_j^{b \theta'}$. These two components are both static,
describing time-averaged quantities. The temporal fluctuations of
the spike train and their possible correlations in time are
captured by the third term on the right-hand side of
(\ref{eq:S-separation}), $\delta S_j^{b \theta'}(t)$. Using the
central limit theorem and methods like those in
\cite{FulviMari:2000} and \cite{Kree+Zippelius:1987} we can then
derive the following mean-field formulation of the recurrent
current:
\begin{equation}\label{eq:I-rec-meanfield}
  I_{a \theta}^ {\rm rec}(t) = \sum_{b=1}^2 J_{ab}
    \left( \sqrt{K_b} A_b + \sqrt{1-K_b/N_b} B_b(t) \right),
\end{equation}
with
\begin{eqnarray}
  \label{eq:A_b}
  A_b &=& \frac{1}{n} \sum_{\theta' = \theta_1}^{\theta_{n}}
    (1 + \gamma \cos 2(\theta - \theta')) r_b(\theta') \\
  \label{eq:B_b}
  B_b(t) &=& \frac{1}{n} \sum_{\theta' = \theta_1}^{\theta_{n}}
    \sqrt{1 + \gamma \cos 2(\theta - \theta')}
    \left(
    \left(\overline{(r_j^{b\theta'})^2}\right)^{\frac{1}{2}} x_{b\theta'} +
    \xi_{b\theta'}(t)
    \right)
\end{eqnarray}
where the values $x_{b\theta'}$ are drawn from a unit-variance
normal distribution. Selecting specific values $x_{b\theta'}$
effectively samples different neurons within the column
population. We have dropped the neuron index $i$ because this
statistical description of the input current reduces the network
problem to single neuron problems -- one for each column
population, indexed by $a\theta$. The terms $\xi_{b\theta'}(t)$
stand for realizations of Gaussian random processes obeying
\begin{equation}\label{eq:xi}
    \langle \xi_{b\theta'}(t) \xi_{b\theta'}(t')
    \rangle =
    C_{b\theta'}(t-t').
\end{equation}
Here, $C_{b\theta'}(t-t')$ denotes the average autocorrelation
function of the fluctuations in the spike trains of neurons with
PO $\theta'$ in population $b$, given by
\begin{equation}\label{eq:C-b}
    C_{b\theta'}(t-t') = \frac{1}{N_b/n}\sum_{j=1}^{N_b/n}
    \langle \delta S_j^{b \theta'}(t) \delta S_j^{b \theta'}(t')
    \rangle.
\end{equation}
With the operation $\langle \cdot \rangle$ we mean averaging over
realizations of random processes, such as stochastic spike trains.
We will refer to such realizations as ``trials" since they
represent (responses to) repeated presentations of the same
stimulus in experimental settings.

\subsection*{The balance condition}

The input currents from the excitatory population and the
inhibitory population have mean values of order $\sqrt{K_1} \gg 1$
and $\sqrt{K_2} \gg 1$, respectively (see
Equation~(\ref{eq:I-rec-meanfield})). In addition, for the
external input current (\ref{eq:I-ext}) we take $I_a^{\rm ext} =
\sqrt{K_0} \hat{I}_a^{\rm ext}$ with $\sqrt{K_0} \gg 1$. If the
neurons are to exhibit irregular firing at a low rate, as cortical
neurons do, these currents must nearly cancel and threshold
crossings have to be caused by the fluctuations in the currents,
which are of order 1. For our orientation hypercolumn model, this
balance condition implies that the average input currents in
(\ref{eq:model}) have to add up to zero for each orientation
column $\theta$:
\begin{equation}\label{eq:balance-condition}
    \sqrt{K_0} \hat{I}_a^{\rm ext}(1 + \epsilon \cos 2(\theta - \theta_0)) +
    \sum_{b = 1}^2 J_{ab} \sqrt{K_b} A_b = {\cal O}(1),
\end{equation}
where $A_b$ is defined in (\ref{eq:A_b}). Here, we have ignored
the contribution of the leakage current (the first term on the
right-hand side of (\ref{eq:model})), because it is small compared
to the input currents, and because the balance condition
(\ref{eq:balance-condition}) holds only up to corrections of
${\cal O}(1)$.

To solve these equations, we consider a continuum formulation for
the weighted average over all angles instead of the discrete
formulation in (\ref{eq:A_b}) and write
\begin{equation}\label{eq:A_b-cont}
    A_b = \int_{-\pi/2}^{\pi/2} \frac{\d \theta'}{\pi}
    (1 + \gamma \cos 2(\theta - \theta')) r_b(\theta').
\end{equation}
Then (\ref{eq:balance-condition}) becomes a pair of integral
equations for $r_a(\theta)$.

In the \emph{broadly tuned case} (all orientation columns respond
with non-vanishing mean rates to every stimulus orientation),
these integral equations can be solved directly. To do so, we
perform a Fourier expansion centered at $\theta_0$ of the mean
rate within orientation column $\theta'$ and write $r_b(\theta') =
r_{b,0} + r_{b,2} \cos 2 (\theta'-\theta_0) + \cdots$. For both
the input current and the connection probabilities, we have
already used such Fourier notations with the fewest possible terms
to retain a periodic function with period $\pi$. Due to that
choice, all higher Fourier components for the mean currents vanish
as well, and we get
\begin{equation}\label{eq:solution-balance-broad}
\fl    \sqrt{K_0} \hat{I}_a^{\rm ext}(1 + \epsilon \cos 2(\theta -
\theta_0)) +
    \sum_{b = 1}^2 \sqrt{K_b} J_{ab}
    [r_{b,0} + \frac{1}{2} \gamma r_{b,2} \cos 2 (\theta-\theta_0)] = 0.
\end{equation}
By solving for each of the two Fourier components of the mean
rates separately, we obtain
\begin{eqnarray}
  \label{eq:mean-rate-F1}
  r_{a,0} &=& - \sum_{b=1}^2 ({\sf \hat{J}}^{-1})_{ab} \hat{I}_b^{\rm ext} \\
  \label{eq:mean-rate-F2}
  r_{a,2} &=& - \frac{2 \epsilon}{\gamma} \sum_{b=1}^2 ({\sf \hat{J}}^{-1})_{ab}
              \hat{I}_b^{\rm ext} = \frac{2 \epsilon}{\gamma} r_{a,0},
\end{eqnarray}
where the matrix ${\sf \hat{J}}$ is composed of the elements
$\hat{J}_{ab} = J_{ab}\sqrt{K_b/K_0}$. Firing rates have to be
non-negative, so this solution can only be valid for $\epsilon \in
(0,\gamma/2]$. However, such a broad tuning is not normally
observed for cortical neurons. Rather, orientation sensitive
neurons tend to be more ``narrowly tuned", with firing suppressed
for stimulus orientations $\theta_0$ that differ too much from the
neuron's preferred orientation $\theta$: $r_a = 0$ for $|\theta -
\theta_0| \ge \theta_c$ for some tuning width $\theta_c$. Within
the parameter regime $\epsilon \in (\gamma/2, \gamma]$ we find
such narrowly tuned solutions to our model. The tuning width
$\theta_c$ turns out to be the same for both excitatory and
inhibitory neurons, which is a consequence of the
population-independence of the tuning parameters $\epsilon$ and
$\gamma$.

To find the solutions for the \emph{narrowly tuned case}, we use
our insight from the broadly tuned case and make the \emph{ansatz}
\begin{equation}\label{eq:ansatz1}
    r_b(\theta') =
    \left\{%
\begin{array}{ll}
    r_{b,0} + r_{b,2} \cos 2(\theta'-\theta_0) & \hbox{for $|\theta'-\theta_0|<\theta_c^b$} \\
    0 & \hbox{for $|\theta'-\theta_0| \ge \theta_c^b$}, \\
\end{array}%
\right.
\end{equation}
where $\theta_c^b = -1/2 \cos^{-1}(r_{b,0}/r_{b,2})$. As mentioned
above, since we have assumed equal tuning in (\ref{eq:ConnProb}),
$\theta_c^b$ is the same for both $b$. Thus, in
(\ref{eq:A_b-cont}) the integration is restricted to
$|\theta'-\theta_0| < \theta_c$. Because $r_b(\theta') = 0$ at
$\theta' - \theta_0 = \theta_c$, we can rewrite the part of the
\emph{ansatz} for $|\theta'-\theta_0|<\theta_c$ in the form
\begin{equation}\label{eq:ansatz}
    r_b(\theta') = r_{b,2}(\cos 2(\theta'-\theta_0) - \cos
    2\theta_c).
\end{equation}
With this approach, we can indeed find solutions for the tuning
width and the rates from the balance condition
(\ref{eq:balance-condition}). Analogous to the solution for the
broadly tuned case (\ref{eq:solution-balance-broad}), now the
total mean-input current can be expressed as
\begin{eqnarray}
\eqalign{
\fl  \langle I^{a \theta, \rm tot} \rangle = \sqrt{K_0} \hat{I}_a^{\rm ext}(1 + \epsilon \cos 2(\theta - \theta_0))\\
    + \sum_{b = 1}^2 \sqrt{K_b} J_{ab}
    [r_{b,2}f_0(\theta_c) + \gamma r_{b,2} f_2(\theta_c) \cos 2 (\theta-\theta_0)],
}
\label{eq:solution-balance-narrow}
\end{eqnarray}
where
\begin{eqnarray}
  \label{eq:f-0}
  f_0(\theta_c) &=& \int_{-\theta_c}^{\theta_c}
   \frac{\d \theta'}{\pi}(\cos 2\theta'-\cos 2\theta_c) =
   \frac{1}{\pi} (\sin 2\theta_c - 2\theta_c \cos 2\theta_c)\\
  f_2(\theta_c) &=& \int_{-\theta_c}^{\theta_c}
    \frac{\d \theta'}{\pi} \cos 2\theta'(\cos 2\theta' - \cos
     2\theta_c) =
    \frac{1}{\pi} (\theta_c - \frac{1}{4} \sin 4\theta_c).
\end{eqnarray}
(We have borrowed the notation from Ben-Yishai \emph{et
al.}~\cite{Ben-Yishai+LevBar-Or+Sompolinsky:1995} who studied a
different kind of model that contains similar expressions; see
also \cite{Hansel+Sompolinsky:1998}). Again, the total current
(\ref{eq:solution-balance-narrow}) has to vanish for all
orientation columns $\theta$, so both the constant and the $\cos
2(\theta-\theta_0)$ terms vanish separately:
\begin{eqnarray}
  \label{eq:mean-rate-F1-narrow}
  \hat{I}_a^{\rm ext} + \sum_{b=1}^2 \hat{J}_{ab} r_{b,2} f_0(\theta_c) &=& 0 \\
  \label{eq:mean-rate-F2-narrow}
  \epsilon \hat{I}_a^{\rm ext} + \gamma \sum_{b=1}^2 \hat{J}_{ab} r_{b,2} f_2(\theta_c) &=&
  0
\end{eqnarray}
Dividing (\ref{eq:mean-rate-F1-narrow}) by
(\ref{eq:mean-rate-F2-narrow}) yields
\begin{equation}\label{eq:theta_c-narrow}
    \frac{f_2(\theta_c)}{f_0(\theta_c)} =
    \frac{\epsilon}{\gamma},
\end{equation}
which can be solved for $\theta_c$. Note that
(\ref{eq:theta_c-narrow}), and thus the tuning width of the mean
rates, does not depend on the overall strength of the input,
$I_a^{\rm ext}$ (i.e., the ``contrast" of the stimulus). We find
therefore contrast-invariant tuning of the mean rates as a result
of cortical interactions, in agreement with experimental findings
\cite{Sclar+Freeman:1982}. Having calculated $\theta_c$, we can
find the mean rates with help of (\ref{eq:mean-rate-F1-narrow}),
via
\begin{equation}\label{eq:r2-narrow}
    r_{a,2} = -\frac{1}{f_0(\theta_c)} \sum_{b=1}^2 ({\sf \hat{J}}^{-1})_{ab}
    \hat{I}_b^{\rm ext},
\end{equation}
and by using the equality $r_{a,0} = -r_{a,2}\cos 2 \theta_c$.

The above calculations show how cortical interactions are
responsible for a narrowing of the tuning of the population firing
rates, relative to the tuning of the input to the network. We can
proceed one step further in our analytical treatment of the
mean-field model and consider the \emph{tuning of the neuronal
input noise spectrum}. We can write the dynamic noise in the input
current as
\begin{equation}\label{eq:input-noise}
    \langle \delta I_{a\theta}^{\rm rec}(t) \delta I_{a\theta}^{\rm rec}(t') \rangle
    = \sum_{b=1}^2 J_{ab}^2
      \int_{-\pi/2}^{\pi/2} \frac{\d \theta'}{\pi}
      (1 + \gamma \cos 2(\theta - \theta')) C_{b\theta'}(t-t'),
\end{equation}
where we have used the continuum notation for the weighted
averages. The correlation function $C_{b\theta'}(t-t')$ has a
piece proportional to $r_b(\theta) \delta (t-t')$, which gives
\begin{eqnarray}
  \lim_{\omega \rightarrow \infty}
      \langle |\delta I_{a\theta}^{\rm rec}(\omega)|^2 \rangle &=&
      \sum_{b=1}^2 J_{ab}^2
      \int_{-\pi/2}^{\pi/2} \frac{\d \theta'}{\pi}
      (1 + \gamma \cos 2(\theta - \theta')) r_b(\theta')
  \\
    &=& \sum_{b = 1}^2 J_{ab}^2
    [r_{b,2}f_0(\theta_c) + \gamma r_{b,2} f_2(\theta_c) \cos 2
    (\theta-\theta_0)].
    \label{eq:noise-sp2}
\end{eqnarray}
To obtain (\ref{eq:noise-sp2}), we performed calculations
analogous to the ones for solving the integrals for the rate
equations. Using (\ref{eq:theta_c-narrow}) and
(\ref{eq:r2-narrow}), we can then write the flat contribution to
the noise spectrum as
\begin{equation}\label{eq:noise-spectrum}
    \lim_{\omega \rightarrow \infty}
      \langle |\delta I_{a\theta}^{\rm rec}(\omega)|^2 \rangle =
    -\hat{I}_a^{\rm ext} [1 + \epsilon \cos 2(\theta-\theta_0)]
    \sum_{b=1}^2 J_{ab}^2
    \sum_{c=1}^2
      ({\sf \hat{J}}^{-1})_{bc} \hat{I}_c^{\rm ext},
\end{equation}
This result states that the high-frequency limit of the neuronal
input noise has the same orientation tuning as the external input
to the neuron.

For $t \neq t'$, it is not possible to calculate analytically
solutions to (\ref{eq:input-noise}) because the correlation
function $C_{b\theta'}(t-t')$ needs to be evaluated numerically.
Similarly, the tuning of the irregularity in the neuronal firing
(as described by, e.g., the Fano factor) can only be determined by
solving the full mean-field model numerically.

\section{Numerical procedure}

In our simulations, we modeled the orientation hypercolumn as an
assembly of 30 orientation columns, with their preferred
orientations $\theta$ equally spaced between $-\pi/2$ and $\pi/2$
(or between $-90$ and $90$ degrees, as in the figure captions). We
used parameter values corresponding to $N_1 = 8000$ excitatory and
$N_2 = 2000$ inhibitory neurons, and a membrane time constant of
$\tau = 10$~ms for all neurons. The generic intra-cortical
connection strengths $J_{ab}$ in (\ref{eq:Jij}) were taken as
\begin{equation}\label{eq:Jmatrix}
        \left( \begin{array}{cc}
        J_{11} & J_{12} \\
        J_{21} & J_{22}
    \end{array} \right) =
    \left( \begin{array}{cc}
        0.5 & -2 \\
        1   & -2
    \end{array} \right).
\end{equation}
The synaptic strengths of the afferent inputs from the LGN are
taken to be stronger for the excitatory neurons; specifically, in
(\ref{eq:I-ext}), we chose $I_2^{\rm ext} = \case23 I_1^{\rm
ext}$. To study the role of the overall strength of synapses, we
multiply the generic synaptic weights (including the strength of
the external input) by a common scaling factor $J_s$.

We use an iterative approach that was originally developed for
spin glass models \cite{Eisfeller+Opper:1992} to find
self-consistent solutions of the firing statistics given by the
rates $r_a(\theta)$, the rate fluctuations
$\overline{(r_j^{a\theta})^2}$, and the correlations
$C_{a\theta}(t-t')$. We start with initial estimates of these
quantities, which we obtain by using a white-noise approximation
in the analytical treatment described above. We then generate many
realizations of Gaussian synaptic currents using (\ref{eq:I-ext})
and (\ref{eq:I-rec-meanfield}), which we use to drive single
integrate-and-fire neurons. By collecting their firing statistics,
we obtain improved estimates of the rates, rate fluctuations, and
correlations. These are then used to repeat the cycle until the
input and output statistics are consistent.

For the hypercolumn model, we need to determine these firing
statistics for each population $a$ (excitatory and inhibitory)
within each orientation column $\theta$. However, because of the
inherent symmetry in the network topology, at each iteration step
we only need to run simulations for half of the columns and mirror
the results  to obtain improved statistics for the entire network.
To collect the firing statistics from the column population
$a\theta$, we run many trials of single neurons that are sampled
from the entire column population. This is achieved by generating
Gaussian input currents that fluctuate not only in time (by
generating realizations of the dynamic and appropriately colored
input noise $\xi_{b\theta'}(t)$ in (\ref{eq:B_b})), but also
differ in their overall size due to the random numbers
$x_{b\theta'}$ in (\ref{eq:B_b}), which reflects the fact that
different neurons have in general different connectivity patterns.
(Note that we have used here -- as throughout the text -- the
indices $a\theta$ for referring to the ``target column", whereas
$b\theta'$ runs over all ``source columns"). For a more detailed
account on handling some of the subtleties in obtaining the
correct statistics, see \cite{Lerchneretal:2004}.

Once the procedure converges, which takes tens to hundreds of
iterations, depending on the set of parameters and the specific
approach taken, one has obtained a set of self-consistent firing
statistics, describing the population responses for a specific
network input (stimulus contrast and stimulus orientation).
Equipped with these population statistics we can then calculate
input and firing statistics for individual neurons. To specify
such a neuron, we select a set
\begin{equation}
    \label{eq:rand-neuron-set}
    \{x_{b\theta'}: b = 1,2;\  \theta' = \theta_1,\ldots,\theta_n\}
\end{equation}
and keep it fixed over all trials to collect the statistics for
that neuron. The $x_{b\theta'}$ represent the intrinsic
variability across the population in the strength of synaptic
input due to the randomness in the connectivity of the network.

\section{Results}

We concentrate first on results describing response
characteristics of neurons obtained from their firing statistics.
It is possible to compare these results directly with known
properties like contrast-invariant tuning, or with the variability
in spike counts. We then describe results pertaining to properties
of the neuronal input currents (and their orientation tunings) for
the hypercolumn model.

\subsection{Tuning of the neuronal firing}

\begin{figure}[t]
\centering
 \includegraphics[width=10cm]{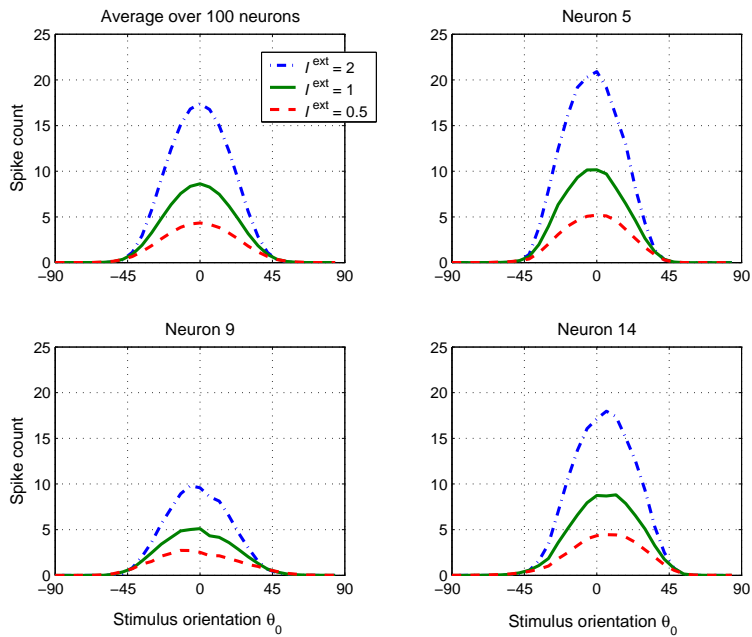}
 \caption{Contrast-invariant tuning width. Average over 100 neurons (upper-left panel)
 and three randomly chosen neurons. The parameter values for the stimulus
 tuning and the connectivity tuning were $\epsilon = 0.5$ and
 $\gamma = 0.625$, respectively, resulting in a tuning width of 43.2 degrees according
 to the calculations. Contrast-invariant tuning is observed for both averaged
 and single-neuron tuning, despite the small distortions and asymmetries
 for single neurons. (See the text for further details)}
 \label{fig:contrast-invariance}
\end{figure}

For the present model, we have shown analytically above that the
tuning width of the column population rates is invariant with
respect to the contrast of the stimulus (see
Equation~(\ref{eq:theta_c-narrow})). We investigated whether such
contrast-invariant tuning is also observed for single, randomly
chosen neurons. The number of afferent connections that a given
neuron receives from neurons with another preferred orientation is
a random number drawn from a probability distribution given by
(\ref{eq:ConnProb}). This will in general distort the shape of the
neuron's tuning curve. Figure~\ref{fig:contrast-invariance} shows
the tuning curves of three randomly chosen neurons from the column
with $\theta = 0$ for three different contrasts $\hat{I}_a^{\rm
ext} = 0.5, 1$, and 2. For our network with 30 orientation columns
and 2 populations, the resulting realization of the random
connectivity to a single neuron is therefore determined by a set
of 60 random numbers (see Equation (\ref{eq:rand-neuron-set})). To
record the neuronal responses, these sets were held fixed, while
the network was presented successively with stimuli of all
orientations $\theta_0$. Also shown in
Figure~\ref{fig:contrast-invariance} is the result of averaging
over the tuning curves of $n = 100$ randomly chosen neurons. While
the averaged tuning is both smooth and symmetric, the tuning
curves of single neurons show small distortions and asymmetries.
Additionally, the overall strength of the response varies from
neuron to neuron. However, despite the somewhat irregular shapes,
the contrast-invariance of the tuning width is preserved for
single, randomly chosen neurons. The analytical treatment predicts
a threshold-cosine shape of the tuning, while the curves shown
here, including the averaged ones, show a rounded fall-off to zero
with non-zero rates for angles just outside the tuning width. This
``rounding artifact" appears to be due to a slow convergence of
the numerical procedure at extremely low firing rates; the
artifact is reduced when the algorithm is run for more iterations.

In all our simulations, we observe an almost linear input-output
relationship between stimulus contrast and firing rate, in
agreement with experiments (see, e.g., Figure~1 in
\cite{Sompolinsky+Shapley:1997}).
\begin{figure}[t]
\centering
 \includegraphics[width=10cm]{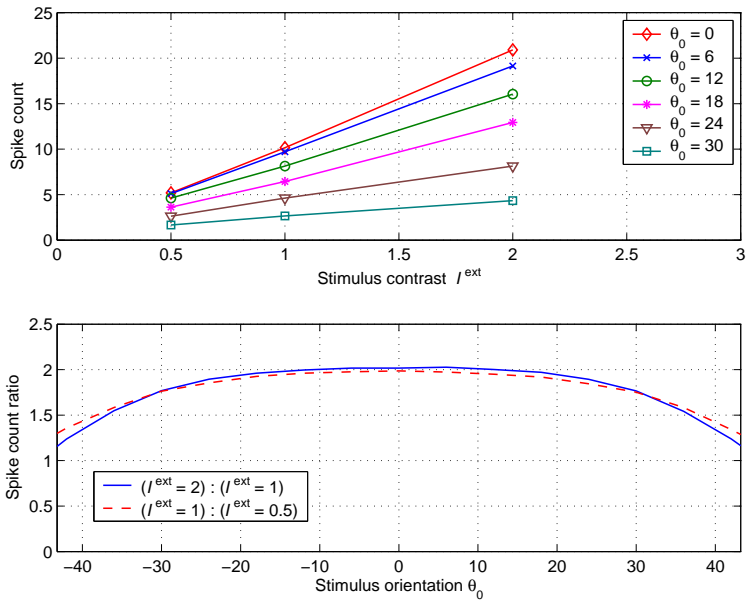}
\caption{Tuning of the gain function. \emph{Upper panel: }Spike
count as a function of stimulus contrast, parameterized by the
stimulus orientation. The input-output relationship is linear, and
the slope decreases as the stimulus orientations $\theta_0$
becomes more dissimilar to the neuron's preferred orientation
$\theta = 0$ (results shown for neuron 5 in
Figure~\ref{fig:contrast-invariance}). \emph{Lower panel: } Spike
count ratios for two pairs of spike counts resulting from doubling
the contrast. At the preferred orientation (PO) and for
orientations not too far from the PO, doubling the contrast
doubles the spike count. For more dissimilar stimulus
orientations, the ratios decrease systematically.}
 \label{fig:Linear}
\end{figure}
Figure~\ref{fig:Linear} shows how the input-output relationship
depends on the stimulus orientation. In the upper panel of
Figure~\ref{fig:Linear}, the spike count is plotted as a function
of the external input strength $\hat{I}_1^{\rm ext}$, i.e. the
contrast of the stimulus, for a single neuron (neuron~5 of
Figure~\ref{fig:contrast-invariance}). The slope changes
systematically with stimulus orientation $\theta_0$, getting
smaller as the difference between the stimulus orientation and the
neuron's preferred orientation increases. The lower panel of
Figure~\ref{fig:Linear} shows the spike count ratios of two pairs
of spike counts that resulted from doubling the stimulus contrast.
In contrast to the upper panel of Figure~\ref{fig:Linear}, these
curves show results of averaging spike counts over 100 neurons, in
order to make the general tendency clearer. It can be seen that
for the preferred orientation, doubling the stimulus almost
perfectly doubles the spike count (this is also true for single
neurons, as can be read off from
Figure~\ref{fig:contrast-invariance}). This relationship also
holds for stimulus orientations away from the PO, until about 20
degrees difference, which is about half the tuning width of these
neurons. For larger orientation differences, the ratio decreases.
It seems likely that at large orientation differences (near the
tuning width) this reduction is due to the rounding artifact for
very low spike rates discussed above. For intermediate orientation
differences, say 20--35 degrees, the reason for the reduction is
not evident to us.

\begin{figure}[t]
\centering
 \includegraphics[width=10cm]{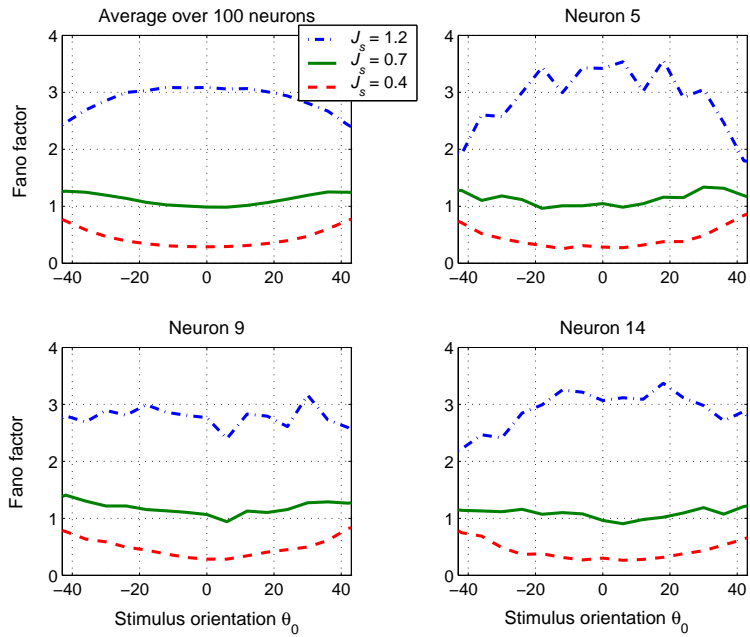}
\caption{Tuning of the Fano factors. Tuning curves, parameterized
by relative synaptic strengths $J_s$, are shown for the same three
neurons as in Figure~\ref{fig:contrast-invariance} and for an
average over 100 neurons (upper left panel). The Fano factors $F$
depend systematically on $J_s$: stronger synapses lead to higher
Fano factors. On average, $F$ stays either above 1 for all
orientations or below 1 for all orientations. For $F \approx 1$,
the tuning is almost flat, while it reaches a maximum (resp.
minimum) at the preferred orientation for $F > 1$ (resp. $F <
1$).}
 \label{fig:Fano-tuning}
\end{figure}

We characterize the irregularity in the neuronal firing by the
Fano factor $F$. For a Poisson process $F = 1$, while $F \neq 1$
implies temporal correlations in the spike times: $F > 1$
indicates a tendency towards ``bursty" spiking behavior, and $F <
1$ indicates more regular spike trains with narrower interspike
interval (ISI) distributions. Figure~\ref{fig:Fano-tuning} shows
the tuning of the Fano factor for three different overall
connection strengths $J_s = 0.4, 0.7$, and $1.2$. As in
Figure~\ref{fig:contrast-invariance}, the results for (the same)
three individual neurons are shown, as well as an averaged tuning
curve. It can be seen that the Fano factor depends systematically
on the overall strength of connectivity: stronger synapses lead to
more irregular spike counts. The averaged tuning curves reveal two
further properties, which we observed consistently in all our
simulations, performed with many different sets of parameters:
First, Fano factors are either less than 1 at all angles or
greater than 1 at all angles. Second, if they are considerably
greater than 1, they peak at the preferred orientation, falling
off as the difference between stimulus orientation and PO
increases; in the case where $F$ stays below 1, the opposite
tuning is observed, i.e., the Fano factor reaches a minimum at the
preferred orientation.
\begin{figure}[t]
\centering
 \includegraphics[width=10cm]{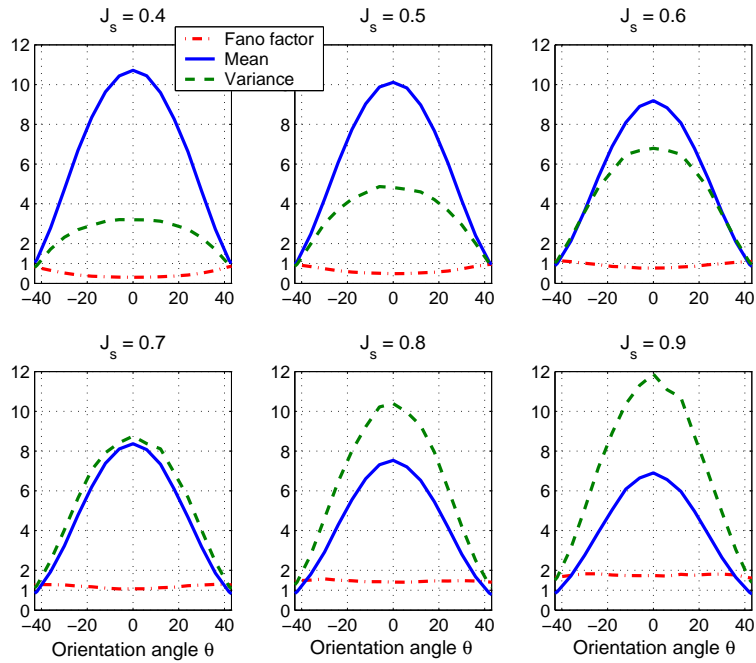}
\caption{Analysis of Fano factor tuning: tuning of mean spike
count and spike count variance for relative synaptic strengths
$J_s = 0.4, 0.5,\ldots,0.9$. For each $J_s$, the variance stays
either below the mean or above the mean for all orientations
(upper and lower panels, respectively), resulting in ratios $F <
1$ and $F > 1$ for all orientations. The variance increases with
$J_s$ -- most sensitively at the preferred orientation (PO). For
$F \approx 1$, the variance and mean tuning curves are almost
identical, resulting in an almost flat tuning of their ratio $F$,
while for $F \neq 1$ the ratios reach a minimum/maximum at the
PO.}
 \label{fig:FanoMeanVar}
\end{figure}
We can shed some light on the emergence of these two properties by
looking at pairs of tuning curves for the spike count variance and
the mean spike count and then systematically changing the
connection strengths. We show these tuning curves for 6 different
values of $J_s$ in Figure~\ref{fig:FanoMeanVar}. It can be seen
that both the mean and the variance peak at the PO, falling off
towards increasing angle differences. Furthermore, for $F \approx
1$ at $J_s = 0.7$, the tuning curves are nearly identical
resulting in almost untuned Fano factors close to 1. For lower
$J_s$ values, the variance curve stays entirely below the mean
curve, while the opposite is true for $J_s$ values bigger than
$0.7$. Therefore, the ratio of the curves, which is the tuning
curve of the Fano factor, stays either always below 1 or always
above 1. The size of the spike count variance depends sensitively
on the overall connection strengths $J_s$. Apparently, this
sensitivity is strongest at the PO, decreasing towards greater
angle differences. Therefore, the Fano factor reaches its minimum
for the cases with $F < 1$ (respectively its maximum for $F > 1$)
when the stimulus is at the preferred orientation.

\begin{figure}[t]
 \centering
 \begin{tabular}{cc}
  \begin{minipage}{6cm}
    \includegraphics[width=5.5cm]{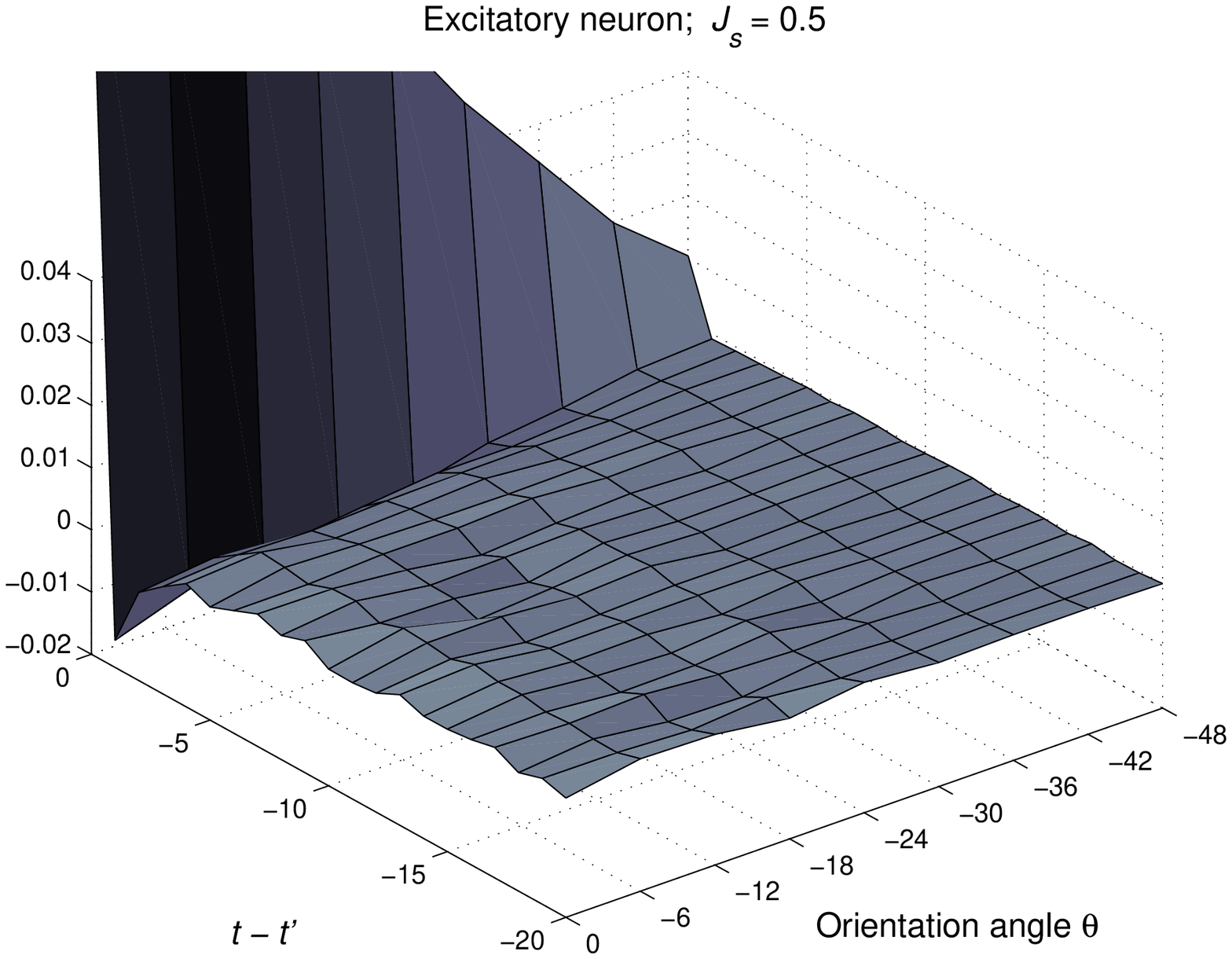}
  \end{minipage}
  \begin{minipage}{6cm}
    \includegraphics[width=5.5cm]{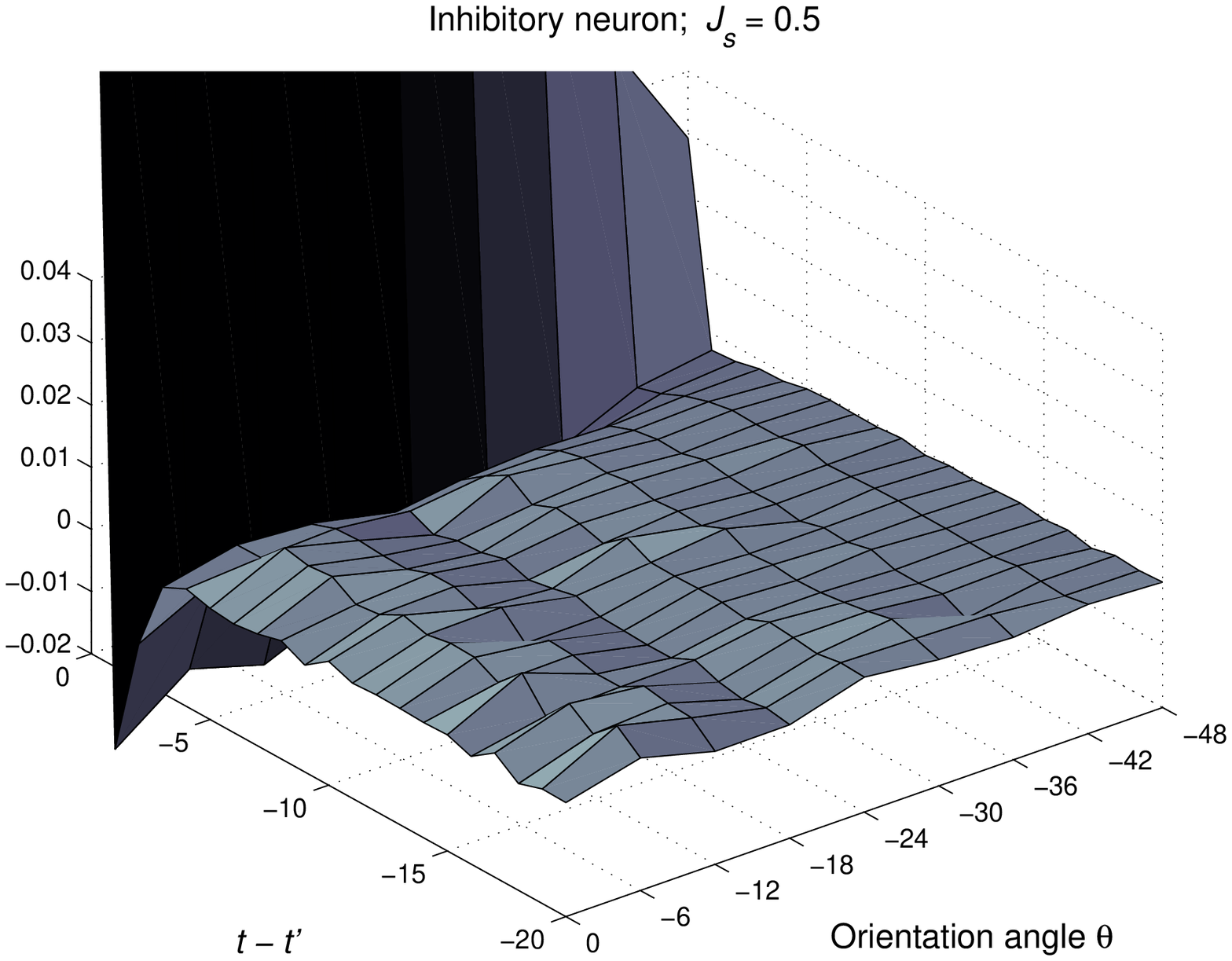}
  \end{minipage}
 \\
  \begin{minipage}{6cm}
    \includegraphics[width=5.5cm]{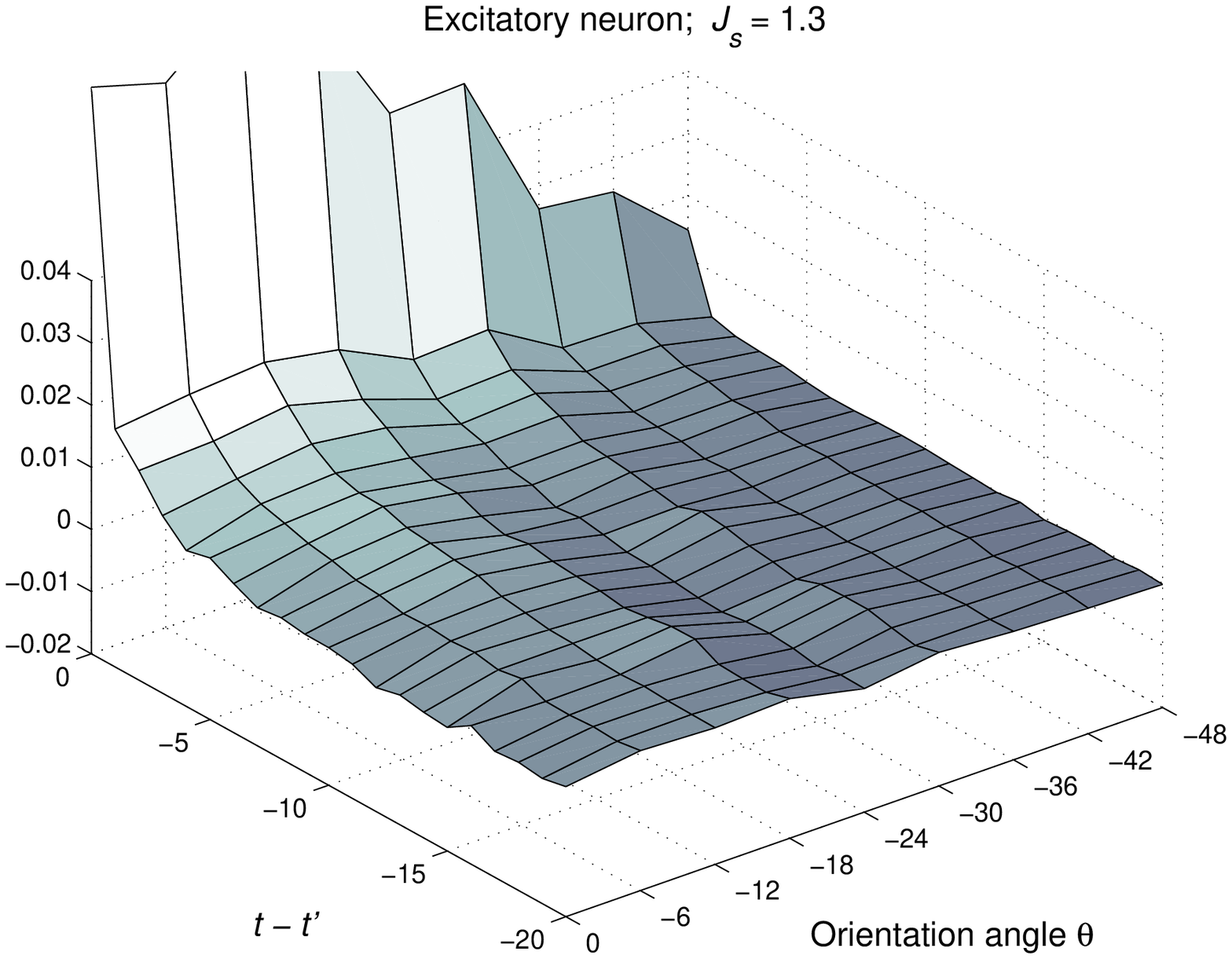}
  \end{minipage}
  \begin{minipage}{6cm}
    \includegraphics[width=5.5cm]{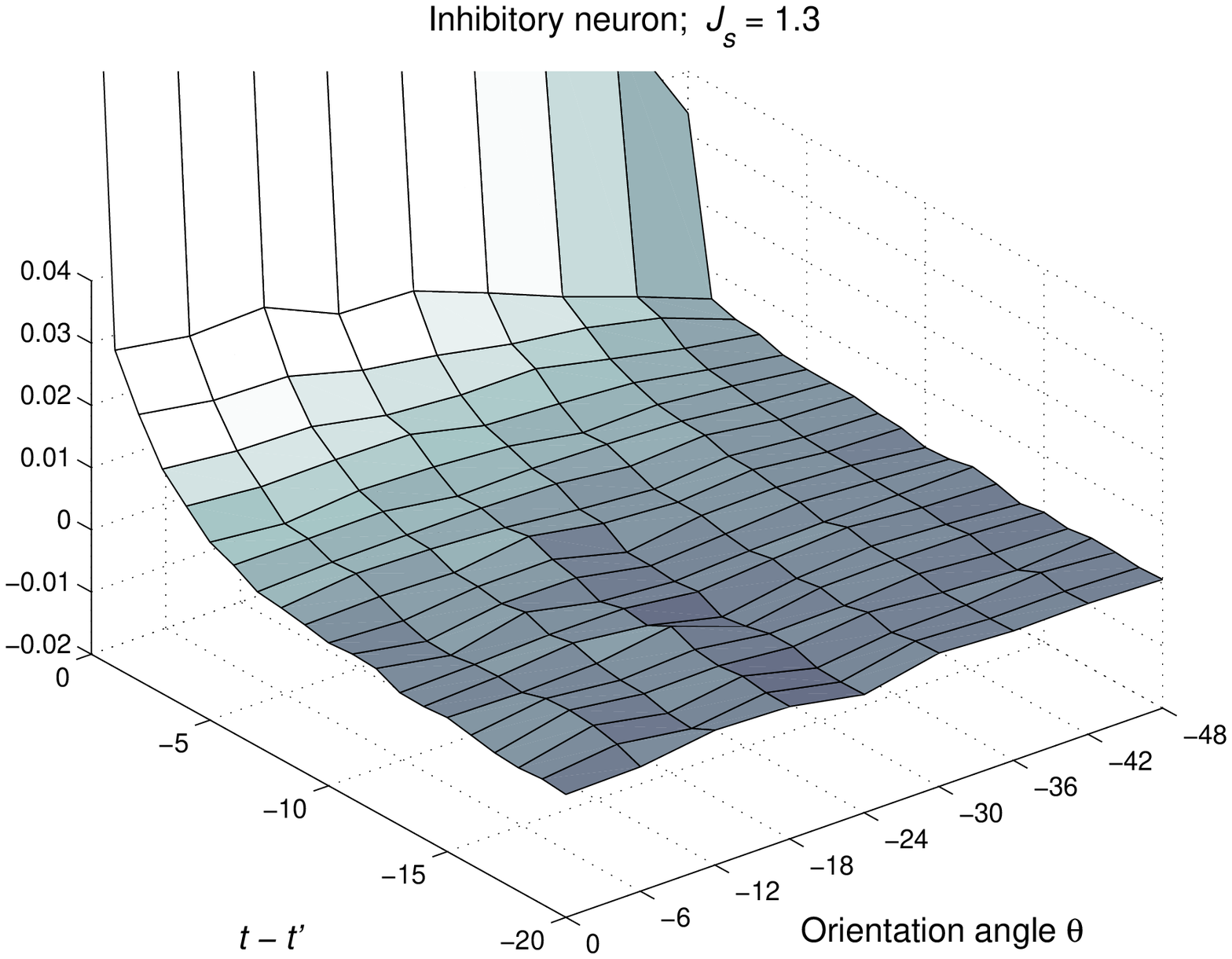}
  \end{minipage}
 \end{tabular}
\caption{Autocorrelation tuning. \emph{Upper panels:} Weak
synapses with $J_s = 0.5$. There is a dip to negative values for
small time differences. It decreases in strength at greater time
differences. The dip indicates a relative refractoriness to
emitting a spike immediately after a previous one, resulting in
Fano factors $F < 1$. \emph{Lower panels:} Strong synapses with
$J_s = 1.3$. There is a hill of positive correlations for short
intervals, falling off to zero for increasing time differences.
The hill indicates a tendency toward clustered spikes, resulting
in $F > 1$. The autocorrelations for excitatory neurons (left
panels) and inhibitory neurons (right panels) show the same
qualitative features, differing only in overall size. }
 \label{fig:Auto-tuning}
\end{figure}

As already mentioned, Fano factors that deviate from 1 indicate
temporal correlations in the spike trains. The nature of these
correlations and their orientation dependence is summarized in
Figure~\ref{fig:Auto-tuning} for a case with $F < 1$ ($J_s = 0.5$;
upper panels) and a case with $F > 1$ ($J_s = 1.3$; lower panels)
for both excitatory neurons (left panels) and inhibitory ones
(right panels). For $J_s = 0.5$, there is a negative dip for small
time differences, indicating a relative refractoriness to emitting
a spike immediately after a previous one. For stronger synapses
($J_s = 1.3$) there is no such refractoriness. On the contrary,
for strong synapses, we observe positive correlations for small
time differences. For both strong and weak synapses, the
correlations are strongest at the preferred orientation and
decrease monotonically for less optimal stimulus orientations. The
autocorrelations for excitatory and inhibitory neurons show the
same qualitative features, differing only in their overall size.

\begin{figure}[t]
\centering
 \includegraphics[width=11cm]{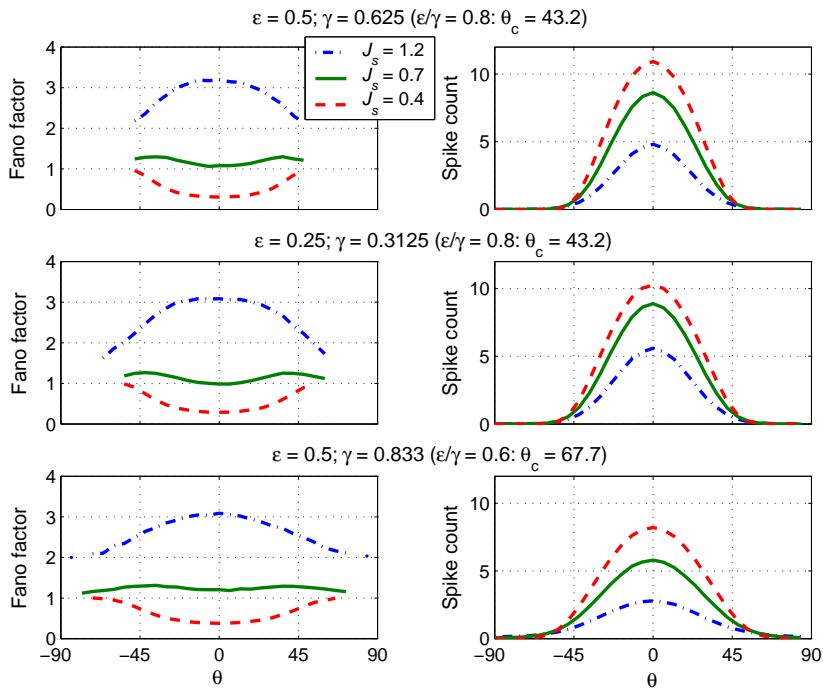}
\caption{Dependence of the Fano factors on tuning parameters
$\epsilon$ and $\gamma$ at three different values of relative
synaptic strengths $J_s$. Fano factors and mean spike counts are
shown for three different combinations of $\epsilon$ (external
input tuning) and $\gamma$ (connectivity tuning). The tuning of
both the Fano factors and the mean counts are controlled by the
ratio $\epsilon/\gamma$.}
 \label{fig:eps-gammaFano}
\end{figure}

In Figure~\ref{fig:eps-gammaFano} we illustrate how the firing
statistics depend on $\epsilon$ and $\gamma$, which determine how
strongly the input current and the intracortical connectivity are
tuned (see equations (\ref{eq:I-ext}) and (\ref{eq:ConnProb}),
respectively). Fano factor tuning curves (left panels) and firing
rate tuning curves (right panels) for three different combinations
of $\epsilon$ and $\gamma$ are shown, parameterized by $J_s$, the
scaling factor for the synaptic strengths. As shown analytically
above, the ratio $\epsilon/\gamma$ determines the tuning width of
the neuronal firing (see Equation~(\ref{eq:theta_c-narrow})). This
is reflected by the identical firing tuning widths in the first
and second row of Figure~\ref{fig:eps-gammaFano}, for both of
which $\epsilon/\gamma = 0.8$, resulting in a tuning width of
$\theta_c = 43.2$ degrees. The third row of
Figure~\ref{fig:eps-gammaFano} shows results for the same external
input tuning $\epsilon = 0.5$ as in the first row, but for a
different ratio $\epsilon/\gamma = 0.6$. This results in $\theta_c
= 67.7$ degrees and an accordingly broader tuning curve of the
firing, plotted in the right panel of the third row. The curves
for the Fano factor tuning in the left panels of
Figure~\ref{fig:eps-gammaFano} suggest that the tuning of the
firing irregularity is -- just as the tuning of the firing itself
-- only dependent on the ratio $\epsilon/\gamma$. (We consistently
found this dependence in all our simulations.)

\subsection{Tuning of the neuronal input current}

\begin{figure}[t]
\centering
 \includegraphics[width=11cm]{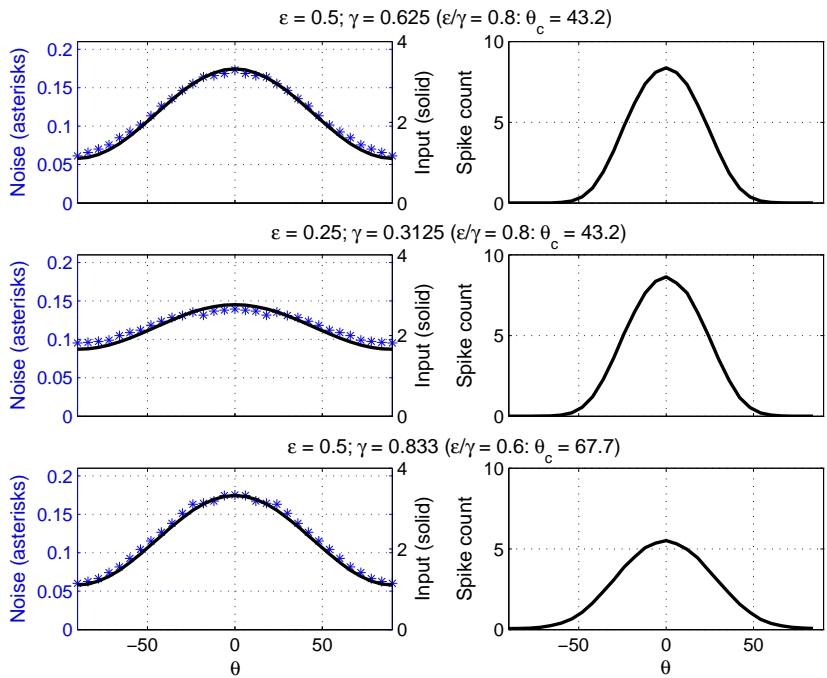}
\caption{Dependence of the noise on tuning factors $\epsilon$ and
$\gamma$. External input and dynamic input noise versus tuning of
the neuronal firing for the same three combinations of $\epsilon$
and $\gamma$ as in Figure~\ref{fig:eps-gammaFano}. It can be seen
that the tuning of the noise is determined by $\epsilon$, while
the tuning of the firing rate is determined by the ratio
$\epsilon/\gamma$.}
 \label{fig:eps-gammaNoise}
\end{figure}

Our analytical treatment of the balanced hypercolumn model reveals
that the high-frequency neuronal input noise power has the same
tuning as the external input. In Figure~\ref{fig:eps-gammaNoise}
we show simulation results of the noise tuning for the same three
combinations of $\epsilon$ and $\gamma$ as in
Figure~\ref{fig:eps-gammaFano}. For the panels in the first and
the second row of Figure~\ref{fig:eps-gammaNoise},
$\epsilon/\gamma = 0.8$, but $\epsilon = 0.5$ and $\epsilon =
0.25$ in the upper and middle rows, respectively. While the tuning
of the neuronal firing is identical for these two cases, the noise
tuning is weaker in the middle row, reflecting the weaker tuning
of the external input (left panels). The results presented in the
third row of Figure~\ref{fig:eps-gammaNoise} show a case with a
broader tuning of the response, resulting from a different ratio
between $\epsilon$ and $\gamma$, but with the same $\epsilon =
0.5$ as in the first row. For these two cases, the tunings on the
input side -- concerning external input and dynamic noise -- are
practically indistinguishable, while the tunings of the firing
differ. Thus, the noise tuning is determined by $\epsilon$, unlike
the response tuning, which depends on the ratio $\epsilon/\gamma$.

\begin{figure}[t]
\centering
 \includegraphics[width=11cm]{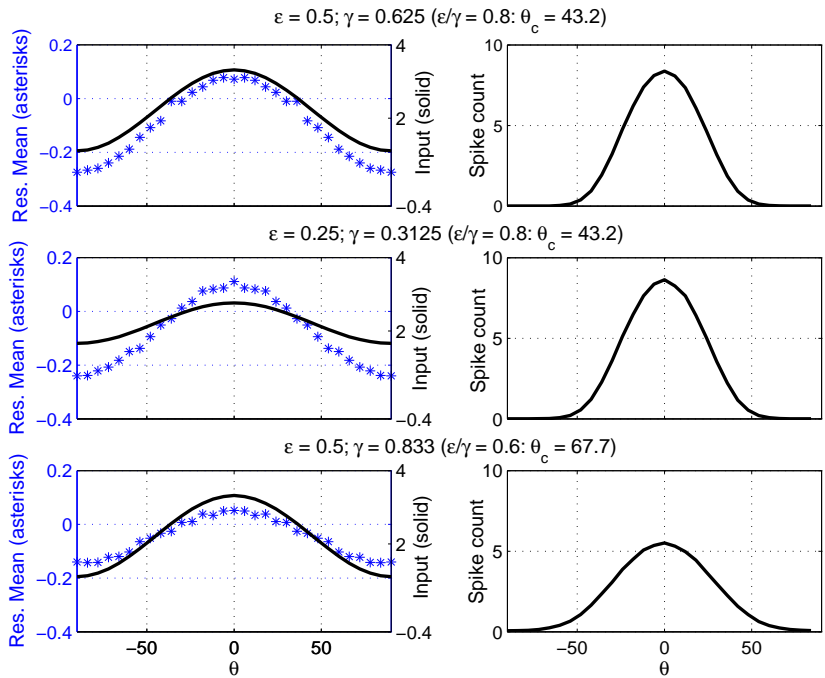}
\caption{Dependence of the mean input current on tuning factors
$\epsilon$ and $\gamma$. External input tuning and mean-input
tuning versus tuning of the response for the same three
combinations of $\epsilon$  and $\gamma$ as in
Figure~\ref{fig:eps-gammaFano} and
Figure~\ref{fig:eps-gammaNoise}. The tuning of the mean input is
not determined by $\epsilon$; rather, as for the spike count
tuning shown in the right panels, the ratio $\epsilon/\gamma$
plays an important role.}
 \label{fig:eps-gammaResmean}
\end{figure}

The balanced state for the orientation hypercolumn implies that
the mean input currents (external and recurrent currents), which
are each of ${\cal O} (\sqrt{K_a})$ with $K_a \gg 1$, cancel up to
corrections of ${\cal O}(1)$. It is not straightforward to
calculate the tuning of the resulting net mean current, since the
balance condition (\ref{eq:balance-condition}) does not allow
inferences about its size. However, the solutions obtained by the
numerical algorithm provide direct access to the net mean
currents, which we depict in Figure~\ref{fig:eps-gammaResmean} for
the same combinations of $\epsilon$ and $\gamma$ as for the noise
tuning in Figure~\ref{fig:eps-gammaNoise}. It is clear from
Figure~\ref{fig:eps-gammaResmean} that the tuning of the mean
input, unlike the dynamic input noise tuning, is not determined by
the tuning of of the external input. Rather, it seems to be the
ratio $\epsilon/\gamma$ that primarily determines it, as suggested
by the almost identical tunings for the two cases with identical
$\epsilon/\gamma$. Since the tuning of the external input and that
of the noise variance are the same, the left panels of
Figure~\ref{fig:eps-gammaResmean} also show how the tuning of the
noise compares to that of the mean input current for the three
combinations of $\epsilon$ and $\gamma$.

\section{Discussion}

In this work, we presented a complete mean field theory for a
balanced network with structural inhomogeneity, together with an
algorithm that allows one to find the self-consistent solutions
for the mean rates, their cell-to-cell fluctuations, and the
correlation functions. We applied the theory to a simple model of
an orientation hypercolumn in primary visual cortex, comprised of
integrate-and-fire neurons. Despite the relative simplicity of the
model, the resulting dynamics capture several key properties known
about responses of orientation selective cortical neurons \emph{in
vivo}. Within this description, we can pinpoint how the resulting
neuronal dynamics are controlled by parameters of the model, and
quantify their influence.

Specifically, we find contrast-invariant tuning of the neuronal
firing not only for the population rates, as derived from the
analytical treatment, but also for single, randomly chosen
neurons. Moreover, the firing rate increases linearly with the
strength of the input current (i.e., the contrast of the
stimulus). Note that these are network effects originating in the
dynamical balance between excitation and inhibition, not
properties of isolated neurons. This is in agreement with
experimental results, where such a linear input-output
relationship can only be found for cortical neurons \emph{in
vivo}, but not for single neurons \emph{in vitro}.

Another network effect that emerges naturally from the
self-consistent dynamic balance, in combination with the static
randomness in the connectivity, is the irregularity in the
neuronal firing. We are able to describe it quantitatively through
the correlation functions, which are determined self-consistently
in the theory. Such firing-statistical issues cannot be addressed
in ``rate models", which simply assume a particular relation
between average input current or membrane potential and firing
rate. While it is possible to calculate the firing variability in
the mean-field treatment of Brunel~\cite{Brunel:2000}, it cannot
be done in a self-consistent manner because of the assumption that
the neuronal input is uncorrelated in time (white noise). Here we
color the noise self-consistently. Poisson-like statistics (Fano
factor $F = 1$) are only one possibility within a continuum of
firing statistics that depend sensitively on the strengths of the
synapses: stronger synapses generally lead to higher Fano factors.
The underlying mechanism can be summarized as follows: Stronger
synapses increase the probability of a spike shortly after reset,
which leads to a higher tendency of spikes occurring in
``clusters", thereby increasing the spike count variance. A
detailed account of this mechanism, involving the dependence of
the membrane potential distribution on the synaptic strength can
be found in \cite{Lerchneretal:2004}, where the analysis was
carried out for a single cortical column.

The mean field theory applied to the present model allows us to
study tuning properties of both the neuronal firing and the
neuronal input and their dependence on network parameters.
Concerning the irregularity of firing, our results suggest that
$F$ stays either above 1 or below 1 for all orientations.
Moreover, the modulation strength of $F$ over angles increases,
relative to the almost untuned case of $F \approx 1$, with
increasing (resp. decreasing) overall values of $F$, reaching a
maximum (resp. a minimum) at the preferred orientation.

Concerning the tuning of the input currents, we find analytically
that the high-frequency input noise power has the same tuning as
the external input to the neuron (which in turn is determined by a
Hubel-Wiesel feed-forward connectivity from the LGN). In our
numerical calculations we observe a close fit between the tuning
of the overall input noise and the one of the external input. This
suggests that the tuning of the external input may be a good
predictor for the noise tuning, and vice versa. In contrast, we
find that the tuning of the mean input current does not reflect
the one of the external input, but is predominantly determined by
the ratio $\epsilon/\gamma$ of the modulation strengths of the
external input and the cortical interactions.

Some of our results (the existence of a stable, asynchronous
low-rate state, contrast-invariant orientation tuning, and the
inverse relation between the sharpness of orientation tuning and
intracortical tuning strength $\gamma$) were obtained previously
by Wolf \emph{et al.}~\cite{Wolf+vVreeswijk+Sompolinsky:2001} in
an extension of van Vreeswijk and Sompolinsky's stochastic binary
model \cite{vVreeswijk+Sompolinsky:1996,
vVreeswijk+Sompolinsky:1998} to a hypercolumn, but the treatment
of a spiking neuron model and all the results for correlations of
both input and output are new here. Also new is that we go beyond
population statistics and make quantitative predictions about
input and output characteristics of \emph{individual} neurons,
which can be tested directly.

Firing irregularity of neurons in primary visual cortex has been
investigated experimentally for a long time (see, e.g.,
\cite{Heggelund+Albus:1978, Dean:1981,
Tolhurst+Movshon+Thompson:1981, Snowden+Treue+Andersen:1992,
Gershon+etal:1998}). Well studied is also the dependence of firing
rate on the stimulus orientation \cite{Sclar+Freeman:1982,
Skottun+etal:1987}, but we are not aware of studies investigating
the dependence of firing irregularity on the orientation. Our
predictions concerning the tuning of the input currents (for both
mean and noise) can be tested experimentally by systematically
changing $\epsilon$ (the external input tuning strength) via
changing the spatial modulation of the stimulus and then observing
how the the mean and noise tunings are affected separately.

The mean field theory presented here, in combination with the
numerical procedure for finding the self-consistent solutions, can
be applied to models that capture more of the known neuronal and
cortical physiology. For example, it is straightforward to
incorporate conductance-based synapses into the hypercolumn model,
as has already been done for a single-column model (see
\cite{Lerchneretal:2004b} and \cite{Hertz+Lerchner+Ahmadi:2004}).
It is also straightforward to use different, possibly more
realistic neuron models -- even several kinds of neuron models
within one given network model, since the neuronal dynamics are
explicitly simulated within the numerical procedure for collecting
the firing statistics. Here, we have shown how the theory can be
applied to networks with non-homogenous architecture, using a
simple one-dimensional model for a cortical hypercolumn. This
model can be thought of as describing an annulus around a pinwheel
center. Using the same general techniques as introduced here, the
model can be extended to incorporate a two-dimensional geometry to
describe an entire pinwheel. Similarly, as we have shown for
orientation selectivity, it is possible to include other coding
features, such as spatial phase, for example. Thus, the power of
this mean-field approach lies in its generality, which makes it
possible to quantify dynamics of balanced, highly connected
networks.

\end{document}